\documentclass[epj,twocolumn]{webofc}
\usepackage[varg]{txfonts}   
\newcommand{\opera}{{\tt OPERA}}

\newcommand{\gm}{\ensuremath{g-2}}
\newcommand{\wa}{\mbox{\ensuremath{\omega_a}}}
\renewcommand{\wp}{\mbox{\ensuremath{\omega_p}}}
\newcommand{\amu}{\ensuremath{a_{\mu}}}

\woctitle{Flavour changing and conserving processes}
\begin{document}
\title{Next Generation Muon $g-2$ Experiments}
%


\author{David W. Hertzog\inst{}\fnsep\thanks{\email{hertzog@uw.edu}}}
\institute{Department of Physics, University of Washington, Seattle, WA, 98195, USA
}

\abstract{%
  I report on the progress of two new muon anomalous magnetic moment experiments, which are in advanced design and construction phases.  The goal of Fermilab E989 is to reduce the experimental uncertainty of $a_\mu$ from Brookhaven E821 by a factor of 4; that is, $\delta a_\mu \sim 16 \times 10^{-11}$, a relative uncertainty of 140~ppb.  The method follows the same magic-momentum storage ring concept used at BNL, and pioneered previously at CERN, but muon beam preparation, storage ring internal hardware, field measuring equipment, and detector and electronics systems are all new or upgraded significantly.  In contrast, J-PARC E34 will employ a novel approach based on injection of an ultra-cold, low-energy, muon beam injected into a small, but highly uniform magnet.  Only a small magnetic focusing field is needed to maintain  storage, which distinguishes it from CERN, BNL and Fermilab.  E34 aims to roughly match the previous BNL precision in their Phase~1 installation.}
\maketitle
\section{Introduction}
\label{intro}
At this Workshop, more than 25 presentations were devoted to topics centered on the muon anomalous magnetic moment, $a_\mu \equiv (g-2)/2$. These included a discussion of the two new measurement campaigns -- as reported in this paper -- and the theoretical issues related to both the Standard Model prediction and various new-physics speculations. In the narrative below, I will focus mainly on the Fermilab E989 experiment, which is in a mature construction state.  An important update since this Workshop has been the successful commissioning of the superconducting storage ring magnet to its design field value and start of field-shimming operations. I will also briefly summarize the unique approach to measuring $a_\mu$ being taken by J-PARC E34 and point out, where it is important, the key differences between the experiments, \cite{Gorringe:2015cma}.

\subsection{The $g-2$ test}
Muon $g-2$ is a special quantity because it can be both measured and predicted to sub-ppm precision, enabling the so-called $g-2$ test for new physics defined by $a_\mu^{\rm New} \equiv a_\mu^{\rm Exp} - a_\mu^{\rm SM}$.  As a flavor- and CP-conserving, chirality-flipping, and loop-induced quantity, $a_\mu$ is especially sensitive to new physics contributions~\cite{Czarnecki:2001pv,StockingerFCCP}.
The current $g-2$ test gives:
\begin{equation*}
\Delta a_{\mu}^{\rm New} = [(263-289) \pm 80] \times 10^{-11}  ~~~ (3.3-3.6)\,\sigma.
\label{eq:deltaamu}
\end{equation*}
The range here represents different, but standard, evaluations~\cite{Davier:2010nc,Hagiwara:2011af} of hadronic vacuum polarization (HVP) loops, and a common averaged value for hadronic light-by-light (HLbL) scattering.  In fact, the range is even wider if all efforts to evaluate the SM are considered.  These topics received extensive scrutiny at this Workshop.  If this range for $\Delta a_{\mu}^{\rm New}$ is confirmed at a greater significance, the positive sign and relatively large magnitude -- several times greater than the electroweak contribution -- will provide important clues to the physics it is trying to reveal.  The key phrase above is, ``at a greater significance.'' Let's first focus on the combined experimental and theoretical uncertainty of $80 \times 10^{-11}$ and investigate what contributes to this value and how the combined error might be reduced in the future.

\subsection{The Standard Model Inputs}
The SM terms are usually listed in five categories:
\begin{equation*}
a_\mu^{{\rm SM}} = a_\mu^{{\rm QED}}
 + a_\mu^{{\rm Weak}} + a_\mu^{{\rm HVP}} + a_\mu^{{\rm Had-HO}} + a_\mu^{{\rm HLbL}}
\end{equation*}
with a quadrature summed total uncertainty of $\sim50 \times 10^{-11}$.

The QED and Weak term uncertainties are totally negligible.  These are impressive calculations. At this Workshop, Melnikov provided an interesting discussion exploring in particular the QED calculations and whether they could be wrong in any way; he strongly argues ``no''~\cite{MelnikovFCCP}.

The HVP contribution~\cite{JegerlehnerFCCP} is determined from experiment through a dispersion relation that amounts to an energy-weighted integral of $e^+ e^- \rightarrow$~{\em hadron} total cross sections.  The uncertainty at $42 \times 10^{-11}$ is non-negligible and presently dominates the overall $\delta a_\mu({\rm SM})$.  This contribution depends on the accuracy of the reported data. Its quoted uncertainty is arrived at from reported cross section errors and, when necessary, errors are expanded to account for independent data sets that differ beyond statistical expectations. A small theoretical uncertainty owing to radiative corrections is also important.  New experimental campaigns at BESIII~\cite{Ablikim:2015orh} and the upgraded VEPP-2000 facility~\cite{EidelmanFCCP} in Novosibirsk, along with continued analyses of the large and varied BaBar data set~\cite{ZhangFCCP} can be counted on to reduce the HVP uncertainty.

Higher-order HVP diagrams contribute a value of $-98.4 \times 10^{-11}$ to $a_\mu({\rm SM})$, with negligible uncertainty.  On the other hand, the higher-order HLbL effect is more problematic. Curiously, it has a total contribution of approximately $+105 \times 10^{-11}$, essentially canceling the HVP-HO term.  At present, it can only be estimated using hadronic models, which typically have various strengths, weaknesses, and different limitations.  Assigning an uncertainty is almost a guess.  We use $26 \times 10^{-11}$, which is a consensus value reached by comparing models, but one could as well chose a number 50\% larger, which many people do. More troubling is that these models could be badly wrong.

Fortunately intense efforts using lattice QCD have been making rapid progress toward a prediction of HLbL with an uncertainty goal below 10\%.  Lehner described recent efforts by the RBC and UKQCD Collaborations that aim to give 10\% - 20\% uncertainty soon on the quark-connected diagrams~\cite{LehnerFCCP,Blum:2015gfa}. Work in progress is focussed on the more difficult disconnected diagram, but the message given is that the end goal is achievable; it will be a matter of statistics more so than method.

An additional HLbL thrust is aimed at developing a data-driven approach using dispersion relations for the photon-photon scattering amplitude. Promising efforts described here by Procura~\cite{ProcuraFCCP} will make use of experimental spectra from BES~III and KLOE that are now being harvested.
\subsection{The Experimental Inputs}
The measurement of $a_\mu$ is based on the following principles.  When a muon with charge $q$ is circulating in the horizontal plane of a magnetic storage ring, its cyclotron frequency is $\vec{\omega}_c = -q\vec{B}/ m \gamma$.  The muon spin precesses at frequency $\vec{\omega}_s = -(gq\vec{B} / 2 m) - [(1-\gamma) q\vec{B} /\gamma m]$, owing to the torque on the magnetic moment and including the Thomas precession effect for the rotating reference frame~\cite{Bargmann:1959gz}.  The magnitude of $\omega_s$ is greater than $\omega_c$ for $g \neq 2$.  The difference is the anomalous precession frequency defined by
\begin{equation}
\vec{\omega}_a \equiv \vec{\omega}_s - \vec{\omega}_c = -\left( \frac {g-2}{2} \right)
\frac{q\vec{B}}{m} = -a_\mu \frac{q\vec{B}}{m},
\label{eq:simpleomega}
\end{equation}
where we have assumed for now a negligible effect from a non-zero electric dipole moment.

Parity violation in $\mu^+ \rightarrow e^+{\bar{\nu}}_{\mu}\nu_e$ associates the decay positron energies in the laboratory frame with the average spin direction of the muon at the time of the decay, such that the highest-energy positrons are preferentially emitted when the muon spin is aligned with its momentum and lower-energy positrons are emitted when the spin is reversed.  Systems of detectors measure the decay positron times and energies.

To achieve the conditions described above, polarized muon bunches are injected into the magnet, kicked onto a stable storage orbit, and are then observed non-intrusively until they decay.
In practice, a $>95\%$ polarized muon beam can be obtained by capturing forward decays of in-flight pions $(\pi \rightarrow \mu \nu)$ in a beamline lattice made of alternatively focussing then defocussing (FODO) quadrupole magnets.  The captured beam will have a small, but finite, fractional transverse momentum component such that, unaided, this beam could not be stored in a simple transverse magnetic field; the muons would escape if not for some kind of vertical containment field.  At Fermilab, and earlier at BNL and CERN, this is provided by an electric quadrupole system, which creates a sort of Penning trap.

The motional magnetic field seen by a relativistic muon passing through an electric field $\vec{E}$ contributes an important term to the spin precession rate, represented by
\begin{equation}
   \vec{\omega}_{a} =  -\frac{q}{m}\left[ a_{\mu} \vec{B} -
   \left( a_{\mu}- \frac{1}{\gamma^2 - 1} \right)
   \frac{\vec{\beta}  \times \vec{E}}{c}\right].
   \label{eq:omega}
\end{equation}
At a muon momentum of $p_\mu = 3.094$~GeV/$c$, ($\gamma = 29.4$),
the 2nd term in Eq.~\ref{eq:omega} exactly vanishes.
The residual effect for muons slightly off the magic momentum, and therefore not centered in the null region of the electric quadrupoles, results in an $E$-field correction to the measured precession frequency.
The beam also executes horizontal and vertical betatron motions at frequencies determined by the weak-focussing index of the storage ring (i.e, the electric field strength).  The vertical undulation of the muons means $\vec{p}_\mu$ is not exactly perpendicular to $\vec{B}$, thus a small ``pitch'' correction is necessary.
Combined, these corrections shift $a_\mu$ by $86(6) \times 10^{-11}$~\cite{Bennett:2006fi}; the error was negligible in E821, but will need to be reduced for E989.  This will be accomplished by more sophisticated particle tracking and by indirect measurements of the muon beam profile vs. time using an in-vacuum straw tracker system.

This detail is provided to contrast to the J-PARC experiment, which does not need such corrections.  They will use a muon beam created by re-accelerating an ultra-cold, stopped muon source to a final injection momentum. The beam is predicted to have a relative transverse momentum of $10^{-5}$, a nearly negligible value.

In both experiments, $a_\mu^{\rm Exp}$ is obtained from two independent measurements -- the anomalous precession frequency and the magnetic field -- plus the evaluation of many systematic error categories. Detectors are used to measure the anomalous precession frequency $\omega_a$ and pulsed proton NMR to measure the magnetic field in terms of the proton Larmor precession frequency, $\omega_p$.  Both measurements involve frequencies that are measured using highly stable precision oscillators. It is further necessary to know the muon distribution in the storage ring for the muon population that contributes to the $\omega_a$ data.  This distribution is folded with similarly determined azimuthally averaged magnetic field moments to give the effective magnetic field seen by the muons,  $\tilde{\omega}_p$ below.  Given these experimentally determined quantities, one obtains $a_\mu$ at the precision needed through the relation
\begin{equation}
a_\mu^{\rm Exp} = \frac{g_e}{2}\frac{\omega_a}{\tilde{\omega}_p}\frac{m_\mu}{m_e}\frac{\mu_p}{\mu_e}.
\label{eq:amuexp}
\end{equation}
In this expression, a $g-2$ experiment reports the ratio of the muon precession frequency to the proton precession frequency, $R \equiv \omega_a/\tilde{\omega_p}$, where all systematic errors from the separate uncertainty table entries have been appropriately evaluated and combined in the uncertainty on $R$.  That is the quantity reported by E821~\cite{Bennett:2006fi}.  From different experiments, one obtains the electron $g_e$ factor~\cite{Hanneke:2008tm}, the muon-to-electron mass ratio, and the proton-to-electron magnetic moment ratio,~\cite{CODATA2014}. Compared to the present $80 \times 10^{-11}$ combined uncertainty in the $g-2$ test, these quantities are all known quite well.  Of course, if any of their values change -- e.g., the mass ratio was adjusted after the E821 final paper was published -- one can update $a_\mu^{\rm Exp}$ easily.

\begin{figure}
\centering
\includegraphics[width=\columnwidth]{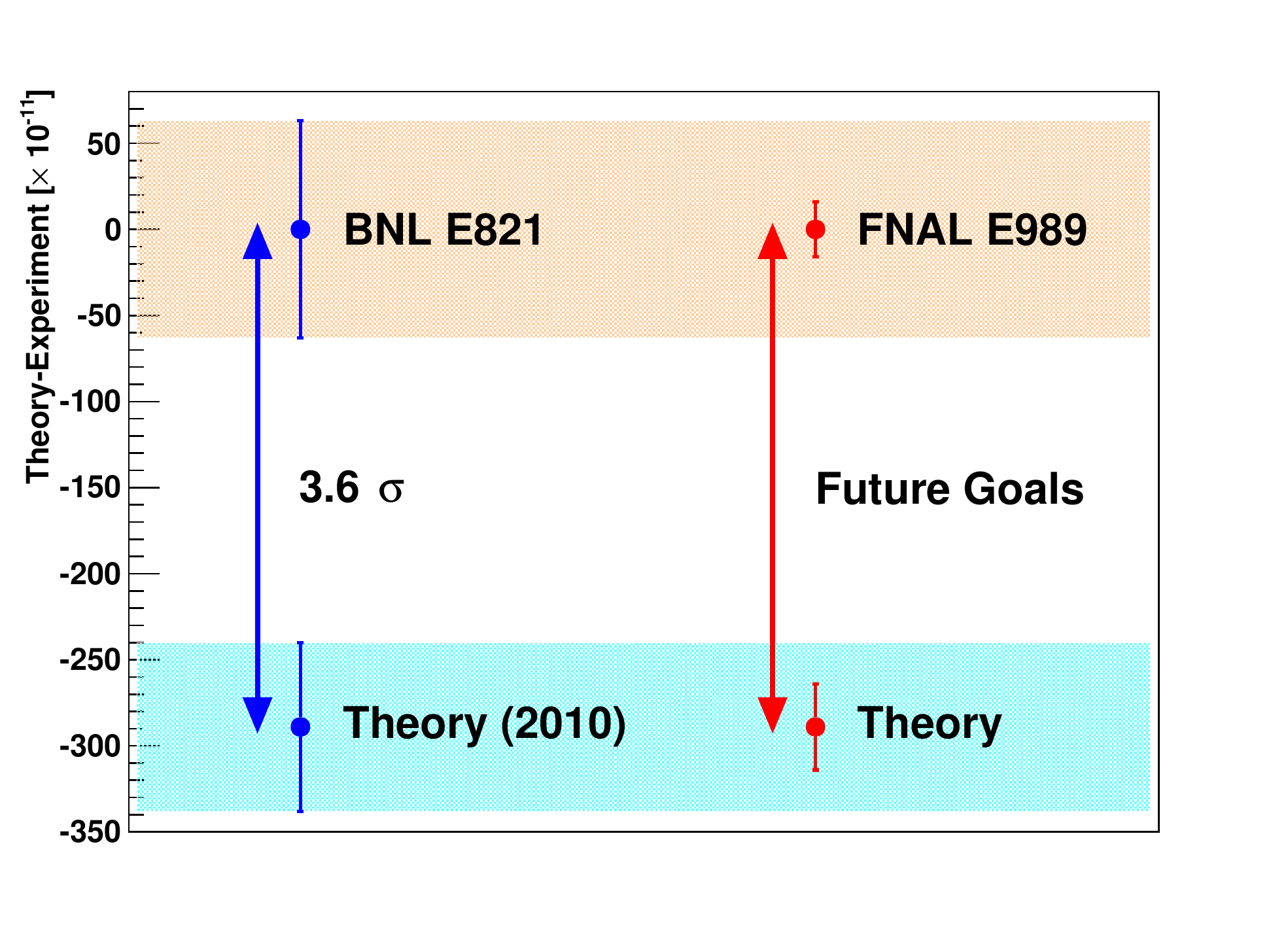}
\caption{Comparison of Experiment to Theory at present and expected on completion of future Fermilab E989 and factor of $\sim 2$ improvements in HVP and HLbL calculations.}
\label{fg:comparison}       
\end{figure}

Table~\ref{tb:errors} summarizes the latest versions of the absolute and relative uncertainties of the theoretical and experimental quantities used in the $g-2$ test.  With the Fermilab E989 goal of $\delta a_\mu \sim 16 \times 10^{-11}$, the quantities that stand out, suggesting targets for improvement, can be identified.  On the experimental side, this will be achieved by a 21-fold increase in statistics and reductions of systematic errors by a factor of 2 to 3.  On the theory side, continued improvement in the HVP evaluation -- perhaps as much as a factor of 2 eventually? -- might be expected.  Furthermore, the promise of recent work on the lattice suggests that a reduction of the HLbL uncertainty to $\sim 13\times 10^{-11}$ is not an unreasonable goal.  If the experiment and theory improvements are met, the uncertainty on the $g-2$ test could reduce from 80 to $30 \times 10^{-11}$ and the New Physics significance for the range quoted in Eq.~\ref{eq:deltaamu} would approach $9 \sigma$.  Figure \ref{fg:comparison} illustrates the $g-2$ test now and in this optimistic future, where the central values are not altered from present values.

\begin{table}
\caption{Uncertainties on the quantities used to determine $a_\mu^{\rm Exp}$ and $a_\mu^{\rm SM}$. Experimental errors from Ref~\cite{Bennett:2006fi}. Theory errors, see Ref~\cite{MelnikovFCCP}, except where noted. CODATA ratio uncertainties from the 2014 online update.}
\label{tb:errors}       
\centering
\begin{tabular}{lcc}
 \hline
  Quantity & Uncertainty & $\delta a_\mu/a_\mu$ \\
           & $\times 10^{-11}$ & (ppb) \\ \hline
  Total $\omega_a$ Statistical  & 53 & 458 \\
  Final $\omega_a$ Systematic & 24 & 210 \\
  Final $\tilde{\omega}_p$ Systematic & 20 & 170 \\
  CODATA $m_\mu/m_e$ & 2.6 & 22 \\
  CODATA $\mu_p/\mu_e$ & 0.35 & 3 \\
  Electron $g$ factor, $g_e$ & 0.000035 & 0.0003 \\ \hline\hline
  QED & 0.08 & 0.7 \\
  Weak & 1 & 8.6 \\
  Had-HO & 0.7 & 6 \\
  HVP (e.g, Ref~\cite{Davier:2010nc}) & 42 & 360 \\
  HLbL  & 26 & 223 \\ \hline\hline
  Net theory  & 49 & 420 \\
  Final E821  & 63 & 540 \\
  Goal Fermilab E989  & 16 & 140 \\
  Goal J-PARC E34 & 47 & 400 \\ \hline
  \hline
\end{tabular}
\end{table}

\section{The Experiments}
\label{sec:exp}
\begin{figure}
  \includegraphics[width=\columnwidth]{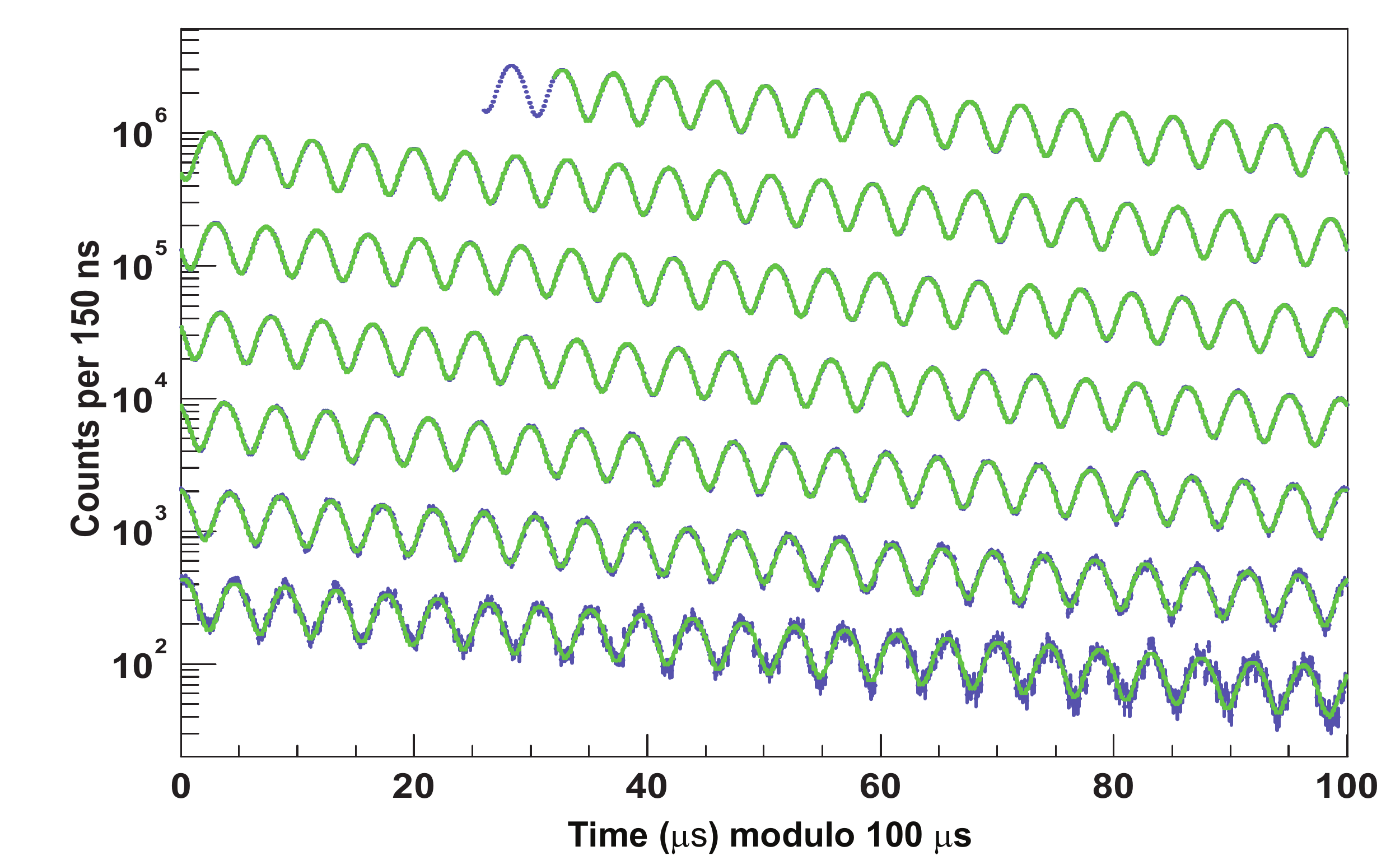}
  \caption{\label{fg:E821data}
    E821 anomalous precession data, including fit.  The data is wrapped around every $100\,\mu$s. Figure courtesy E821 collaboration.}
\end{figure}

The classic data display for the anomalous precession frequency is shown in Fig.~\ref{fg:E821data}.  It represents the arrival time distribution for positrons having energies above threshold $E_{th}$.  This simple method gives a statistical uncertainty on the precession frequency of
\begin{equation}
\delta\wa /\wa = \frac{1}{\wa\gamma\tau_{\mu}}\sqrt{\frac{2}{NA^{2} P^{2}}}.
\label{eq:statistical}
\end{equation}
The uncertainty is improved by running at a higher field ($\omega_a \propto B$), higher momentum ($\gamma\tau$), and high polarization of the incoming muon beam ($P$).  The two experiments being mounted have rather different values for several of these parameters.  In Table~\ref{tb:comparison} -- reproduced from \cite{Gorringe:2015cma} -- the key differences are evident, including the precision goals.  In both cases, the number of events $N$ and the asymmetry $A$ depend on $E_{th}$.  Optimization of the ``figure of merit'' $NA^2$ is achieved for $E_{{\rm th}}/E_{{\rm max}} \approx 0.6$, which both experiments assume.

\begin{table}
    \centering
\caption{Comparison of various parameters for the Fermilab and J-PARC $g-2$ Experiments. Reproduced from \cite{Gorringe:2015cma}.}\label{tb:comparison}
\begin{tabular}{lcc}
  \hline
  Parameter & Fermilab E989 & J-PARC E24 \\ \hline
  Statistical goal & 100\,ppb & 400\,ppb  \\
  Magnetic field & 1.45\,T & 3.0\,T \\
  Radius & 711\,cm & 33.3\,cm \\
  Cyclotron period & 149.1\,ns & 7.4\,ns \\
  Precession frequency, \wa\ & 1.43\,MHz & 2.96\,MHz \\
  Lifetime, $\gamma\tau_\mu$ & $64.4\,\mu$s & $6.6\,\mu$s \\
  Typical asymmetry, $A$ & 0.4 & 0.4 \\
  Beam polarization & 0.97 & 0.50 \\
  Events in final fit & $1.5 \times 10^{11}$ & $8.1 \times 10^{11}$ \\
  \hline
\end{tabular}
\end{table}

\subsection{Fermilab E989}
\label{ssc:E989}

\subsubsection{Beamline}
Fermilab E989 will obtain high-purity bunches of $\sim97\%$ polarized positive muons from a combination of beamlines that are part of the so-called Muon Campus. The sequence is worth describing because it results in an ideal injected beam, having a pion contaminant fraction below $10^{-5}$, and no protons. These hadrons  could have caused a large hadronic ``flash'' at injection, probably paralyzing the detector systems and leading to baseline shifts on a slowly decaying background, as was experienced in E821.  The clean beam is considered to be a major improvement.

The sequence starts at the Booster, where short batches of 8\,GeV protons are delivered to the Recycler Ring.  An RF system separates a batch into 4 tighter bunches of $\sim 10^{12}$ protons each. These bunches are  extracted one at a time and directed to the former antiproton target station, which is tuned to collect +3.1\,GeV/$c$ $\pi^+$.  Pions are transported along a 270\,m FODO lattice and then are injected into the repurposed $\bar{p}$ Delivery Ring (DR), where they make several revolutions before being kicked into a final beamline that terminates at the storage ring entrance. Each bunch is separated by 11~ms or more, and 16 bunches are expected every 1.33-second accelerator cycle leading to an average storage ring fill rate of 12\,Hz. The described beam path involves many changes compared to existing equipment.  While the target system can be used nearly as is, the long beamlines and the DR needs significant work.  A complete end-to-end simulation of the beam, from production to arrival at the storage ring was used to optimize the lattice and to predict the intensity and phase space distribution of muons at injection.  We expect up to $7 \times 10^{5} \mu^+$ per bunch in a $\Delta p/p \approx 2\%$ momentum bite -- which is much greater than the ring acceptance of less than $0.2\%$.

\subsubsection{Storage Ring}
The BNL E821 storage ring~\cite{Danby:2001eh} will be reused at Fermilab.  Following a highly publicized move mostly on water from New York to Illinois of the superconducting coils, and a less-publicized move of hundreds of tons of steel yoke and other components by truck, the storage ring has now been fully re-assembled. It resides in the new MC-1 building, which has a strong supportive flooring system and a uniform and stable year-round temperature control.

During summer 2015, the magnet was first powered, followed by minor repairs to a superconducting lead. Full power and full field were achieved in September.  No muon would be stored in the magnet without a number of critical storage ring subsystems, many of which are being upgraded significantly for E989. Muons first enter the ring through a bored-out tunnel through the back-leg iron yoke of the C-shaped magnet. They pass through the narrow horizontal constrictions of a superconducting inflector magnet, which creates a nearly field-free corridor.  Emerging from the inflector, they traverse a high-gradient field region and cross through the outer electrode Al plate of one of the four electric quadrupole systems before entering the region of uniform magnetic field. One quarter of the way around the ring, a fast magnetic kicker applies an outward transverse 10-11~mrad angular kick during the first turn (only) placing the muon beam on the desired central orbit. Five collimators placed around the ring define the storage volume of 9\,cm diameter transverse and circular storage volume.

This sequence is non-trivial to optimize. The incoming beamline Twiss parameters, the inflector angle and field magnitude, and the individual kicker magnitudes and firing sequence can all be adjusted.  Modern simulation tools were used to recreate the BNL E821 conditions and consequently led to design improvements for E989. For example, by reducing multiple scattering in the quadrupole plates and high-voltage standoff structures, a significant increase in the stored muon fraction can be obtained.  Similarly, if the present close-ended windings on the BNL inflector magnet were to be removed -- a new design is in development with exactly this feature -- the stored muon fraction will be nearly doubled.

One key device that demanded a major upgrade from the outset is the kicker magnet.  For E821, its LCR-based pulse forming network was too slow and the magnitude of the kick too low.  These led to an non-optimized storage fraction and large coherent betatron motions of the stored beam -- a fact that resulted in a fairly large systematic uncertainty on \wa.  The new kicker will be energized through a  Blumlein triaxial transmission line~\cite{Grange:2015fou}.  The pulse rises, flat-tops, and falls, within the 149\,ns cyclotron period of the ring -- an important fact.  With newly shaped kicker plate geometry, the field magnitude can exceed the predicted maximum, such that a tuning sweep can be performed to optimize the storage fraction while also minimizing the coherent betatron amplitude.  The aim of these systems is to result in storage of about 18,000 muons per fill immediately after injection, an increase compared to BNL of about 2.5.

\subsubsection{Precision Magnetic Field}
\begin{figure}
 \includegraphics[width=\columnwidth]{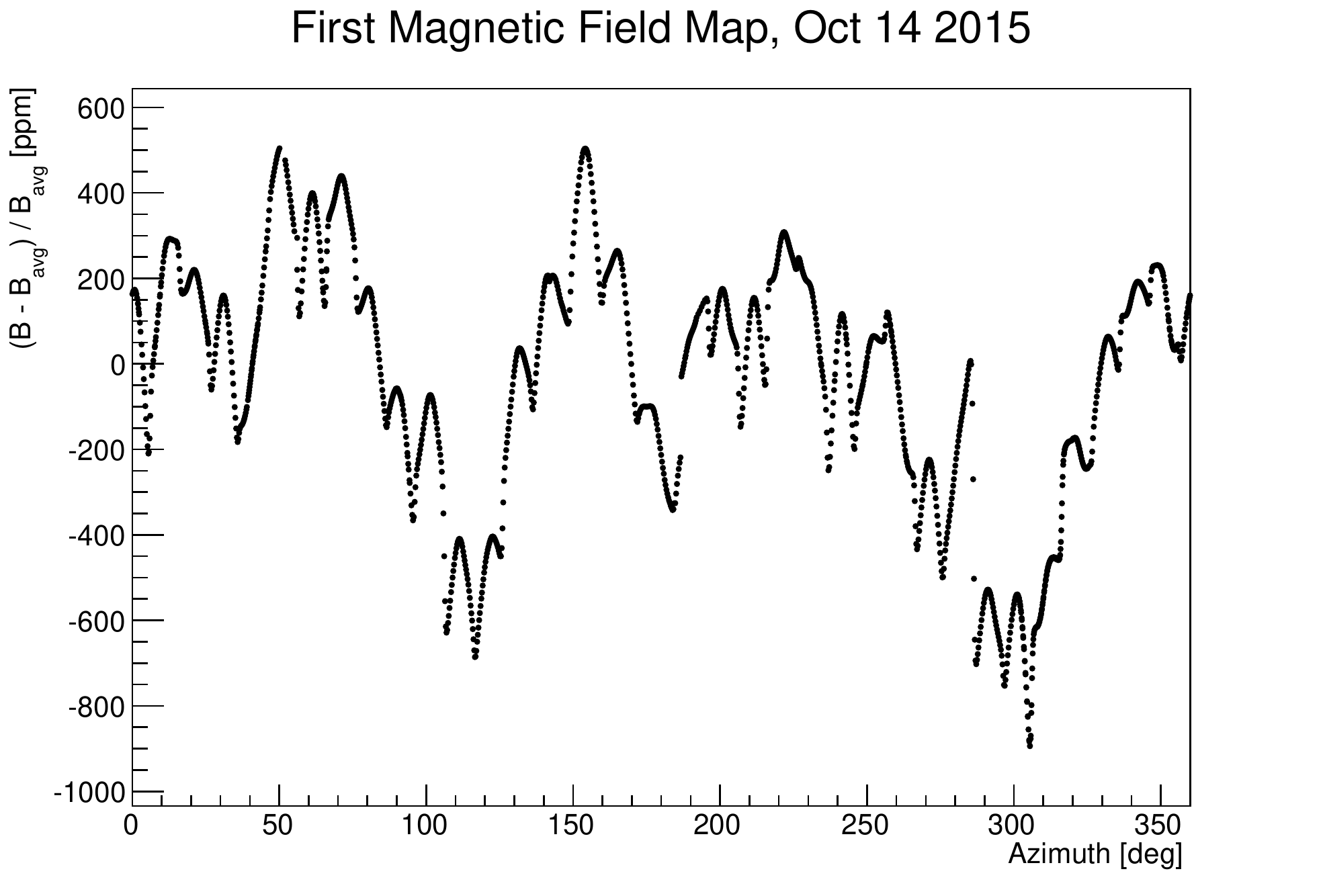}
  \caption{\label{fg:rawfield}
    First ``as built'' raw field map for the reassembled storage ring magnet for E989.  The nominal field is 1.4513 T. The lumpy features will be greatly reduced as the shimming process begins (see text).}
\end{figure}

With the magnet fully powered, the shimming operations designed to smooth out the natural variations in the "as assembled" magnetic field vs. azimuth around the ring have begun.  Although the four energizing superconducting coils are continuous, the surrounding steel is built in coarse units.  Each $30^\circ$ yoke section has three pairs of precision pole pieces, 72 adjustable wedge shims, and sets of adjustable edge shims.  Placement of these components, and the specifics of their dimensions determines the overall field strength in the local region.  The pieces are designed to be adjusted to account for the  natural variance one starts with owing to tiny (10's of microns) placement errors, and intrinsic magnetic material variances. The task began by establishing the baseline ``raw field map'', as shown in Fig.~\ref{fg:rawfield}.  The data is obtained using a specialized shimming trolley having 25 NMR probes and Capacitec sensors that precisely measure the pole gap vs azimuth. The location of the shimming trolley is determined reproducibly by a laser alignment system. The field shown is obviously lumpy on the given scale, which is expected.
The fine humped features stem from the curvature of each pole surface, which leads to a changing pole-to-pole gap and thus a strongly dependent $B$ field.  The longer wavelength features are dominated by three things: the initial conditions of the wedge shims, the top-hat air gap, and the relative pole alignment. Adjustments to flatten these features rely on an iterative process that involves predictive actions based on an \opera\ simulation, followed by re-measurement.

After coarse shimming is completed, the vacuum chamber system will be installed.  It includes the inflector, electric quadrupoles, and the kicker. Further finer adjustments using active current shim coils will then be made.  The field in the storage ring volume will be measured by the {\em in vacuum} trolley with 17 NMR probes, which rides on a set of rails inside the vacuum chamber and records its position using a bar-code reader. It is designed to enter the storage region and map the field exactly where the muons are stored.

All field measurements involve pulsed proton NMR, in which the upgraded data collection procedure will record the free-induction decay (FID) waveforms, rather than simply counting zero crossings.  While the E821 system was indeed excellent, the many intervening years degraded much of the inventory. After evaluating the status of the existing systems, a decision was made to build from scratch the 400 fixed probes and the 17 trolley probes.  The absolute field probe from E821 has been preserved and will be reused in E989. It was also the same probe used in the muonium hyperfine experiment that established the muon-to-proton magnetic moment ratio~\cite{Liu:1999iz}. To complement this key tool, new absolute probes are being developed as further cross checks.  The electronics to provide the $\pi/2$ pulses and to record the FIDs is mostly new, as is much of the internal controller circuitry inside the trolley.

The uncertainty in the field measurement is entirely systematic.  Many of the individual entries in the systematic error table can be reduced naturally by additional, but not radical, efforts.  The field systematic error in E821 was already at the 170\,ppb level; for E989, the goal is 70\,ppb. An underlying improvement that affects many of the systematic issues is the intrinsic field variation in the storage volume -- the goal of the intensive shimming program.  Better location of the trolley during its measuring cycle, reduced trolley temperature fluctuations, greater field stability from improved experimental hall temperature control, a more complete set of fixed NMR probes, and storage of the FID waveforms which allows field determination in regions of higher gradients, are all important subplots to this unfolding story.  As the work has just now begin, we can look forward in about a year to learn how successful the effort has been.

\subsubsection{Detector Systems}
The detector systems for E989 are entirely new, as is the supporting electronics, and the fast and slow data acquisition systems.  The detector systems will include:
\begin{itemize}
  \item An entrance scintillating paddle and several sets of scintillating fiber hodoscopes to determine the incoming muon beam intensity profile vs. time and space.
  \item Two sets of rebuilt scintillating fiber hodoscopes placed in the path of the stored muons to measure (somewhat destructively) the stored beam $x-y$ profile in two locations; these detectors are used to optimize the storage and measure the coherent betatron motion. They are rotated out of the way for normal data taking.
  \item Three sets of in-vacuum straw trackers that reside in the scallop of the vacuum chamber adjacent to the muon storage volume.  They provide data of decay positron tracks that can be ``traced back'' to the point of tangency from the muon that decayed.  This provides a transverse stored-muon profile vs. time-in-fill.  They also serve to calibrate the calorimeters and they are sensitive to an electric dipole moment of the muon (see \cite{ChislettFCCP}).
  \item Twenty-four stations of electromagnetic calorimeters that are positioned to maximally intercept the higher-energy decay positrons and determine their time of arrival and energy.
\end{itemize}

The data that appears in the \wa\ plot shown in Fig.~\ref{fg:E821data} is extracted from calorimeter signals read out using continuous sampling waveform digitizers. The instantaneous rate at BNL approached several MHz per station.  At Fermilab it will exceed that by the 2.5 times higher stored muon rate.  Because pileup was a leading systematic error that scales with rate -- and must be reduced -- a segmented and very fast calorimeter is being designed. Pileup occurs when two decay positrons strike the calorimeter at nearly the same time, and are interpreted as one decay with a larger energy than either individually.  If it exceeds $E_{th}$ such that the event is accepted, the phase will be wrong because the two positrons that led to this fake single ``high'' energy event had each traveled a shorter distance from their parent muons than the single high-energy positron would have that they mimic.  That is, the time from muon decay to hit a detector is shorter on average for the low-energy positrons than the higher-energy positron.  Mitigating the level of intrinsic pileup is accomplished by segmenting the calorimeter (unlike E821) and by choosing a technology that allows two-pulses to be resolved with a separation of just a few nsec.

\begin{figure}
\begin{centering}
  \includegraphics[width=\columnwidth]{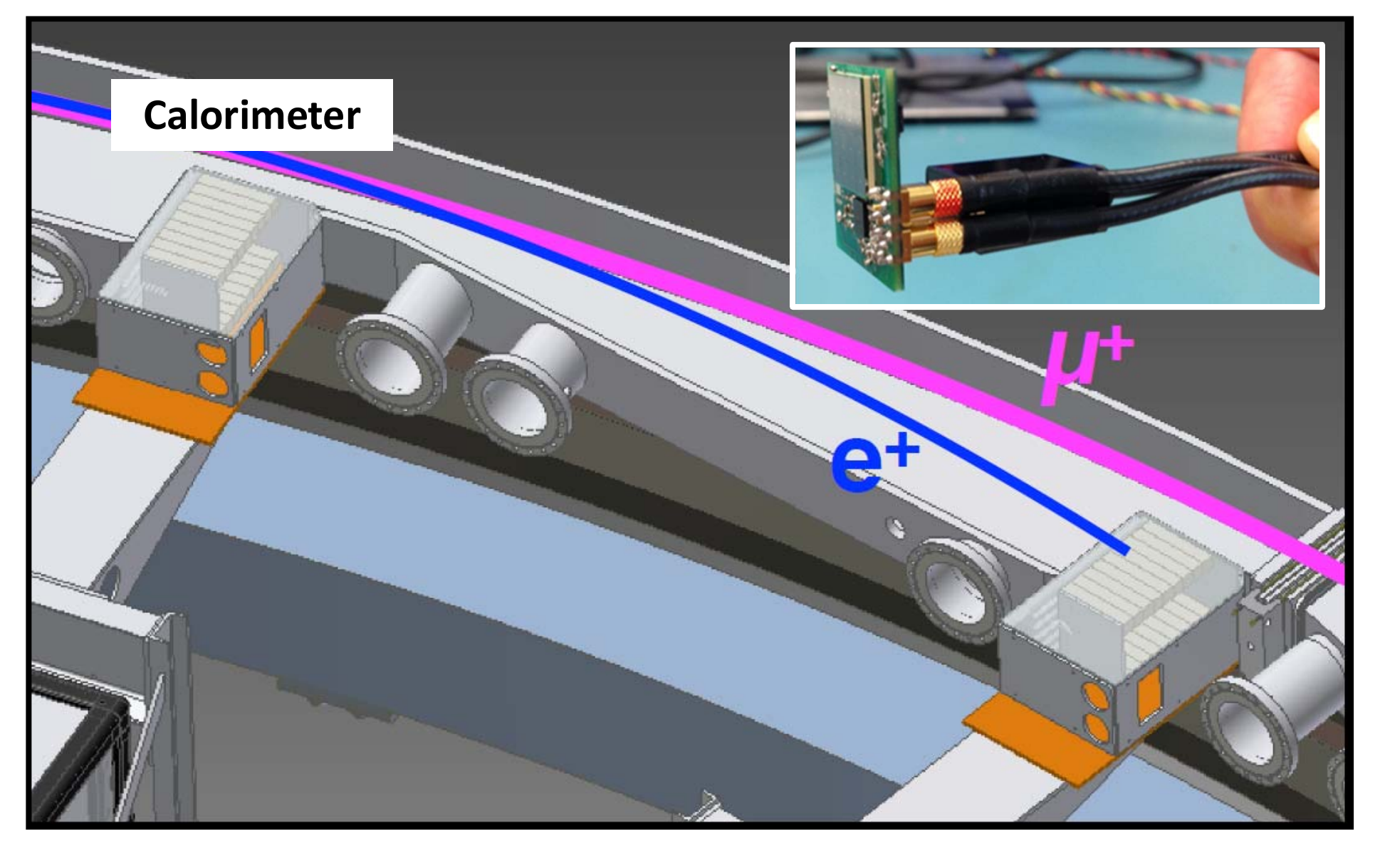}
  \caption{\label{fg:detectors} Positioning of two of the segmented calorimeter stations just downstream of the scalloped section of the vacuum chambers.  The central muon orbit is sketched in, as is a nominal decay positron.  The inset shows the size of the SiPM readout and its custom amplifier and pulse-shaping board.}
\end{centering}
\end{figure}

Each of the E989 calorimeters will consist of 54 PbF$_2$ Cherenkov crystals stacked in a 6 high by 9 wide array. The compact placement, see Fig~\ref{fg:detectors}, and the proximity to the storage ring field, does not allow conventional PMTs to be attached to the downstream face of the crystals. An ultra-fast readout based on the latest generation of Hamamatsu MPPCs (SiPMs, or silicon photomultipliers) will be used; see inset to Fig~\ref{fg:detectors}.  Each SiPM has 57,600 $50\,\mu$m pitch pixels, all operated in Geiger mode.  The summed current is routed to a bank of custom 800\,MHz, 12-bit-depth, waveform digitizers.  The SiPMs have excellent gain stability if the temperature and bias are controlled; they are relatively impervious to rate (tested above 10\,MHz); and, they are not affected by, nor do they affect, the magnetic field.  Continuous digitization of the 54 crystals throughout each fill leads to a raw data rate of 18\,GB/s for the 24 calorimeter stations, which is transferred to a bank of GPU processors where the sparsely occurring individual pulse islands are extracted from the data stream and stored.

Paramount to the systematic control is a very high degree of gain calibration and gain stability, both long term and, even more importantly, short term.  Over a $700~\mu$s fill, the instantaneous rate drops by more than four orders of magnitude, a typical recipe in detectors for a gain change.  To monitor this precisely, a laser calibration system is being developed~\cite{Anastasi:2015ssy} that will flash all crystals prior to each fill and, during special calibration runs, sparsely overlay calibration pulses on top of the real data in a pattern that leads to a series of $10^{-4}$ gain measurements every $5\,\mu$s throughout the fill. The calorimeter system and gain stability were tested at SLAC and the resolution and overall performance was excellent, see~\cite{Fienberg:2014kka}.

\subsubsection{Reduction in Error Summary}
The beamline and improved muon storage instrumentation will lead to the needed $\times 21$ in events to reach the desired statistical error of 100\,ppb. The improved shimming, upgraded NMR system and analysis, and more frequent trolley maps, will reduce the \wp\ systematic uncertainty to a target goal of 70\,ppb. The long and pure beamline will largely reduce the muon loss systematic.  The improved kicker and adjustment of the weak-focussing tune, will reduce the systematic error related to coherent betatron motions.  The $E$-field and pitch corrections are improved by simulation and measurement.  The pileup will be reduced by the segmentation and faster detector response, as well as the many times better resolution and much higher sampling and bit-depth of the digitizers.  Finally, the gain stability will be monitored by the new calibration system.  Overall, the systematic error goal for \wa\ is 70\,ppb. The E989 total error budget is the quadrature sum of the above values, giving $\delta\amu/\amu = 140\,$ppb.

\subsection{J-PARC E34}
\label{ssc:E34}
The J-PARC E34 Collaboration is developing a very unique \gm\ experiment.  They start with production of surface muons -- those born from pions at rest -- which have a momentum of 29\,MeV/$c$.  They are stopped in a silica aerogel target where micro-channels have been created using a laser ablation technique~\cite{Beer:2014ooa}.  A large fraction of muonium M~$\equiv \mu^+e^-$ atoms formed in the target diffuse out and emerge in a near field-free region in vacuum.  The atoms at this point are essentially at rest. The triplet and singlet muonium combinations have relative populations of $\frac{1}{2}, \frac{1}{4}, 0$ and $\frac{1}{4}$, for the $(F,M_F)$ states $(1,1), (1,0), (1,-1)$ and $ (0,0)$, respectively.

Two lasers are fired on the M atoms.  A $\lambda = 122$\,nm shot excites the 1S to 2P transition, and a $\lambda = 355$\,nm shot ionizes the atom, leaving a source of at-rest $\mu^+$. The retained polarization is $P_z = 50\%$, derived exclusively from atoms in the triplet (1,1) state.  The (1,0) and (0,0) populations can make transitions at the rate $\nu_{24} \approx 4.5$\,GHz, which is so high that no net polarized signal will be left.  To date, experiments at TRIUMF have demonstrated~\cite{Beer:2014ooa} a source scaled to J-PARC conditions of about $2 \times 10^5 \mu/$s.

These at-rest, ultra-cold, polarized muons will be accelerated to a momentum of 300\,MeV/$c$ using a new muon linac.  The beam should have negligible transverse dimensions and transverse momentum.  It is, in fact, quite a unique beam that could be used for a $\mu^+ - p$ scattering experiment such as the one proposed by MUSE to help resolve the proton radius puzzle -- but, that is another topic (see \cite{Gorringe:2015cma}).

\begin{figure}
\begin{centering}
  \includegraphics[width=.8\columnwidth]{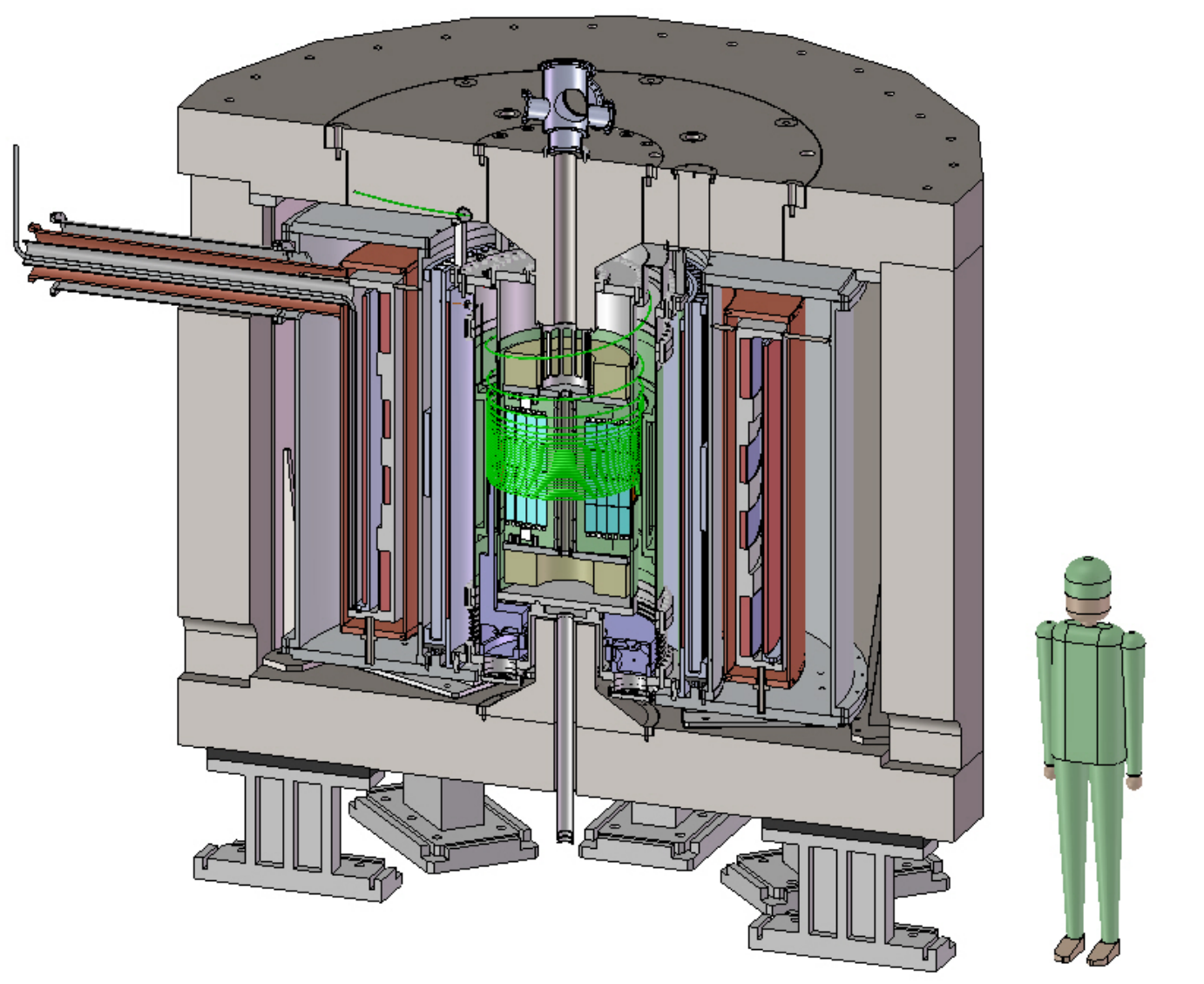}
  \caption{\label{fg:jparcg-2} The proposed setup for the J-PARC \gm\ Experiment.  Muons enter at the top left (green trajectory) and spiral into the highly uniform magnetic field region.  Their decay positrons curl inward to an array of silicon tracking detectors. Figure courtesy T.\,Mibe.}
\end{centering}
\end{figure}

The 10 times lower momentum muons compared to CERN, BNL and Fermilab, implies the need for a much smaller magnet.  Figure~\ref{fg:jparcg-2} shows the compact design, which is based on a conventional MRI magnet, operated at 3\,T. These magnets have rather uniform magnetic field throughout the inner volume.  The muons are injected a steep vertical angle, then they are kicked to the horizontal plane, and finally a small additional magnetic field beyond the solenoid is used to provide vertical containment (it does not appreciably perturb \wa).  Because of the low momentum and high field, the storage radius is just 33\,cm and the cyclotron period is only 7.4\,ns.  The at-most polarization of 50\% results in an asymmetry in the precession data of about half compared to Fermilab for the same relative energy threshold.  Further, the 10 times shorter lifetime reduces considerably the measuring interval for each fill, which affects the precision.

That said, there are many interesting features here. The magnet will not require nearly the attention to establish its mapped uniformity. The detector system uses a radial arrangement of vanes of silicon strip detectors that provide a uniform acceptance for in-curling positrons from any azimuthal decay position.  The rates anticipated are very high, but if the detector system can operate efficiently, it should be relatively free of pileup and perhaps not subject to major gain changes.  The EDM measurement, which both experiments will carry out in parallel, is well suited here as it is based on up/down sloping tracks versus time. The E989 experiment will have only a few tracker stations.

The design of E34 is new enough such that even a working systematic error table has not yet been discussed in public, so it is difficult to predict the final uncertainty they can achieve. The Phase 1 statistical goal is roughly 400\,ppb, a good comparison to BNL E821, but using a very different method.

\section{Summary}
These are exciting times.  As evidenced at this Workshop by the variety of presentations on specific calculations and measurements, many people are working on $g-2$ in one way or another.  The motivation is clear for all.  The current discrepancy between measurement and theory is one of the most promising indications of new physics, and the new experimental campaigns and the continued improvement in theory are leading to ``discovery level'' sensitivity in the $g-2$ test.  Fermilab E989 should begin collecting data sometime in 2017. The collaboration will plan for a first result at roughly the precision level of BNL, but the experiment will run continuously to acquire the much larger data set, well into 2018 and possibly beyond.  As with most modern precision experiments, the analysis is blinded and the emphasis on systematic uncertainty study completion will rule the timing of any result announcements.  We are all looking forward to the next few years of vigorous activity.

\section{Acknowledgments}
The author thanks the organizers of this marvelous Workshop, his many colleagues on the Fermilab E989 $g-2$ experiment, and T.~Mibe and G.~Marshall from J-PARC E34.  The author is supported by the DOE Office of Nuclear Physics, award DE-FG02-97ER41020.

\bibliography{Hertzog}
\end{document}